\documentclass[a4paper,11pt]{article}
\pdfoutput=1 % if your are submitting a pdflatex (i.e. if you have
\usepackage{jcappub} % for details on the use of the package, please
\usepackage[T1]{fontenc} % if needed
\usepackage{subfigure}
\usepackage{parskip}
\usepackage{color}

\title{\boldmath Forecasts of CMB lensing reconstruction of AliCPT-1 from the foreground cleaned polarization data}

\author[a,b]{Jiakang Han,}
\author[a,b,1]{Bin Hu,\note{Corresponding author.}}
\author[c,d,1]{Shamik Ghosh,}
\author[e]{Siyu Li,}
\author[f,g]{Jiazheng Dou,}
\author[d,c]{Jacques Delabrouille,}
\author[e]{Jing Jin,}
\author[e]{Hong Li,}
\author[e]{Yang Liu,}
\author[h]{Mathieu Remazeilles,}
\author[f,g]{Wen Zhao,}
\author[i,j,k]{Pengjie Zhang,}
\author[e]{Zheng-Wei Li,}
\author[e]{Cong-Zhan Liu,}
\author[e]{Yong-jie Zhang,}
\author[l,m]{Chao-Lin Kuo,}
\author[n,o]{Xinmin Zhang}

\affiliation[a]{Institute for Frontier in Astronomy and Astrophysics, Beijing Normal University, Beijing, 102206, People's Republic of China}
\affiliation[b]{Department of Astronomy, Beijing Normal University, Beijing 100875, People's Republic of China}
\affiliation[c]{Lawrence Berkeley National Laboratory, Berkeley, CA 94720, USA}
\affiliation[d]{CNRS-UCB International Research Laboratory, Centre Pierre Bin\'etruy, CPB-IN2P3, Berkeley, IRL2007, USA}
\affiliation[e]{Key Laboratory of Particle Astrophysics, Institute of High Energy Physics, Chinese Academy of Sciences, 19B Yuquan Road, Beijing 100049, People’s Republic of China}
\affiliation[f]{CAS Key Laboratory for Researches in Galaxies and Cosmology, Department of Astronomy, University of Science and Technology of China, Chinese Academy of Sciences, Hefei, Anhui 230026, People's Republic of China}
\affiliation[g]{School of Astronomy and Space Science, University of Science and Technology of China, Hefei 230026, People's Republic of China}
\affiliation[h]{Instituto de Física de Cantabria (CSIC-UC), Avenida de los Castros s/n, 39005 Santander, Spain}
\affiliation[i]{Department of Astronomy, School of Physics and Astronomy, Shanghai Jiao Tong University, Shanghai, 200240, People's Republic of China}
\affiliation[j]{Shanghai Key Laboratory for Particle Physics and Cosmology, Shanghai, 200240, People's Republic of China}
\affiliation[k]{Division of Astronomy and Astrophysics, Tsung-Dao Lee Institute, Shanghai Jiao Tong University, Shanghai, 200240, People's Republic of China}
\affiliation[l]{Department of Physics, Stanford University, Stanford, CA 94305, USA}
\affiliation[m]{Kavli Institute for Particle Astrophysics and Cosmology, Stanford, CA 94305, USA}
\affiliation[n]{Theoretical devision, Institute of High Energy Physics, Chinese Academy of Sciences, 19B Yuquan Road, Beijing 100049, People’s Republic of China}
\affiliation[o]{University of Chinese Academy of Sciences, 19A Yuquan Road, Beijing, People’s Republic of China}

\emailAdd{bhu@bnu.edu.cn}
\emailAdd{shamik@lbl.gov}

\abstract{
Cosmic microwave background radiation (CMB) observations are unavoidably contaminated by emission from various extra-galactic foregrounds, which must be removed to obtain reliable measurements of the cosmological signal. In this paper, we demonstrate CMB lensing reconstruction in AliCPT-1 after foreground removal, combine the two bands of AliCPT-1 (90 and 150~GHz) with Planck HFI bands (100, 143, 217 and 353~GHz) and with the WMAP-K band (23~GHz). 
In order to balance contamination by instrumental noise and foreground residual bias, we adopt the Needlet Internal Linear Combination (NILC) method to clean the E-map and the constrained Internal Linear Combination (cILC) method to clean the B-map. The latter utilizes additional constraints on average frequency scaling of the dust and synchrotron to remove foregrounds at the expense of somewhat noisier maps. Assuming 4 modules observing 1 season from simulation data, the resulting effective residual noise in E- and B-map are roughly $15~\mu{\rm K}\cdot{\rm arcmin}$ and $25~\mu{\rm K}\cdot{\rm arcmin}$, respectively. As a result, the CMB lensing reconstruction signal-to-noise ratio (SNR) from polarization data is about SNR$\,\approx\,$4.5. This lensing reconstruction capability is comparable to that of other stage-III small aperture millimeter CMB telescopes. 
}

\begin{document}
\maketitle
\flushbottom

%%%%%%%%%%%%%%%%%%%%%%%%%%%%%%%%%%%%%%%%%%%
\section{Introduction}
\label{sec:intro}

The main scientific goal of the Ali cosmic microwave background (CMB) Polarization Telescope (AliCPT), a ground-based CMB polarization experiment, is to constrain the primordial gravitational wave signal with high precision in the northern sky \cite{Li:2017drr,Li:2018rwc}. AliCPT-1 is an international collaboration led by the Institute of High Energy Physics in Beijing, with about 100 scientists from China, the United States and Europe. 
%AliCPT-1 is a %China-America collaboration Sino-US joint project leaded by the Institute of High Energy Physics, Chinese Academy of Sciences. 
%The collaboration includes several institutions internationally, such as Stanford University, National Institute of Standards and Technology (NIST), Arizona State University, as well as several Universities from China.
%The telescope integrates two frequencies, namely $90$ and $150$ GHz. The full focal plane can accommodate $19$ modules \cite{2021ITAS...3165289S,2020SPIE11453E..2AS}. Each of the 19 modules contains $1704$ transition edge sensor (TES) detectors, among which half are in $90$GHz and the other half are in $150$GHz. The first stage of AliCPT-1 plans to observe 4000 square degrees of dusty foreground cleaned patch (``deep patch'') in the northern hemisphere for 1 year with $4$ modules, which is dubbed hereafter as ``$\texttt{4 module*yr}$'' stage. According to the currently adopted ``deep patch'' scanning strategy of this stage, the square roots of the harmonic mean of the noise standard variance in temperature are approximately $11~\mu{\rm K}\cdot{\rm arcmin}$ for $90$GHz and $17~\mu{\rm K}\cdot{\rm arcmin}$ for $150$GHz. The full width at half maximum (FWHM) of the beam function are $19$ arcmin for the $90$GHz channel, and $11$ arcmin for the $150$GHz channel.\tblue{ The receiver is currently undergoing the cooling tests in the laboratory at Stanford University.}
The AliCPT-1 telescope is a small/medium aperture telescope designed to carry out microwave measurements in dual bands of 90 and 150~GHz. Its aperture is 72~cm with a 63.6~cm wide focal plane, allowing the design of a receiver that can operate up to 19 transition-edge sensor (TES) arrays, with a total of 32,376 TESes. The full width at half maximum (FWHM) of the beam function for these channels are $19$ arcmin at $90$ GHz and $11$ arcmin at $150$ GHz. The receiver is currently undergoing the integration and cryogenic tests in a laboratory at Stanford University \cite{2021ITAS...3165289S,2020SPIE11453E..2AS}. The first phase of AliCPT-1 is planned as a 1-year observation of a 4000-square-degree dusty foreground clean patch (``deep patch'') in the northern hemisphere using $4$ modules, hereafter referred to as the ``$\texttt{4 module*yr}$'' stage. In order to do forecasting studies with AliCPT-1, we perform end-to-end simulations for a complete observing season. The observed maps based on the adopted `deep-patch' scanning strategy have a depth of about $11~\mu{\rm K}\cdot{\rm arcmin}$ at $90$~GHz and $17~\mu{\rm K}\cdot{\rm arcmin}$ at $150$~GHz. This paper, one of a series of articles on AliCPT-1 scientific forecast studies, focuses on the study of CMB lensing reconstructions of AliCPT-1. 

%Thanks to the aforementioned good noise performance, 
%AliCPT-1 telescope can be used to study other scientific goals besides primordial gravitational wave, including galactic and extra-galactic sciences \cite{Zhang:2021ecp}. Among these secondary scientific cases, CMB lensing is one of the promising subjects.  
The AliCPT-1 telescope can be used for scientific targets other than primordial gravitational waves, including galactic and extra-galactic science \cite{Zhang:2021ecp}. Among these secondary science cases, CMB lensing is one of the promising topics.
According to a previous study \cite{Liu:2022beb} (hereafter, we refer this paper as Liu22), AliCPT-1 can measure the lensing signal with high significance, especially via polarization data. For lensing reconstruction, with 1-year of observation, the Liu22 result shows that the 150 GHz channel is able to measure the lensing signal with $15\sigma$ significance via the quadratic minimum-variance estimator, with the polarization data contributing $6.6\sigma$.  
After 4-year of observation (with ``$\texttt{48 module*yr}$'' data accumulation), the significance can reach $31\sigma$ by including both temperature and polarization data. However, the simulation data used in Liu22 only consider statistical noise but does not consider foreground contamination. 

In order to make our prediction more realistic, in this paper we include galactic and extra-galactic foregrounds in the simulation,  and implement foreground removal methods in the analysis. 
The rest of the paper is structured as the follows. We describe the 
simulation data and foreground cleaning in section \ref{sec:2}. 
In section \ref{sec:3}, we present the reconstructed lensing map, power spectrum and signal-to-noise ratio (SNR). Finally, we conclude in section \ref{sec:4}. 

%For the foregrounds simulation, we adopt a recent version \textcolor{red}{(version?)} of the Planck Sky Model (PSM), based on the original version described in \cite{Delabrouille:2012ye}. 
%As for the foreground cleaning, all the current methods rely on the multi-frequency observations. Hence, in addition to the 90 and 150GHz bands of AliCPT-1, we create `re-observed' map sets using sky maps from Planck HFI four bands \cite{Planck:2013wmz,Planck:2018gnk} and WMAP-K band \cite{WMAP:2003ivt}. The noise maps of Planck HFI are obtained from Planck Archive Legacy, and those of WMAP-K band are generated with its noise covariance matrix. We re-observe the coadded sky map consisting of signal and noise in each band, using the same number of detectors as AliCPT-1’s single band. The detailed pipeline are presented in \cite{Ghosh:2022mje}.  

%%%%%%%%%%%%%%%%%%%%%%%%%%%%%%%%%%%%%%%%%%%
\section{Mock data and foreground cleaning}
\label{sec:2}
To test the performance of the lensing reconstruction, we simulate 199 sets of observational sky maps containing the CMB, foreground radiation, point sources, and instrument noise, in seven frequency bands including the 95~GHz and 150~GHz dual bands of AliCPT-1, the 100~GHz, 143~GHz, 217~GHz, 353~GHz HFI bands of Planck, and the WMAP-K band (23GHz).
We used the cosmological parameters from the Planck 2018 results \cite{Planck:2018vyg} to generate lensed CMB maps with $N_{\mathrm{side}}$ of 1024 using the publicly available software class and \texttt{lenspix} \cite{Lewis:2005tp}\footnote{\url{https://cosmologist.info/lenspix/}}. 
We use the Planck Sky Model (\texttt{PSM})\footnote{A developing version \texttt{V2.3.0}.}, described in \cite{Delabrouille:2012ye}, to simulate the required foregrounds in the data. 
The diffuse galactic emission includes thermal dust, synchrotron, free-free, and spinning dust. We also include 
extragalactic emission includes the SZ effects, the diffuse cosmic infrared background, as well as a population of faint radio sources. For simplicity, we exclude all sources above the Planck detection threshold, with the assumption that those will be excised in a preprocessing stage, leaving only faint residuals at a level that can be safely neglected. 
Model parameters of the foreground emissions are set to be slightly different from the current best-fit values, to avoid any confirmation bias during the foreground removal operation. 
In addition, the distribution and intensity of the point sources and the diffuse foreground map at small scales are derived from simulations rather than templates.
Sky maps are convolved with Gaussian beams to match the angular resolution of each of the frequency bands, and the AliCPT-1's single observational footprint mask is  applied to all observations. The noise maps of Planck HFI are obtained from Planck Legacy Archive \cite{PLA}, and those of WMAP-K band are generated with its noise covariance matrix provided with the frequency maps of WMAP 9-year data \cite{WMAP9}. We re-observe the coadded sky map consisting of signal and noise in each band, using the same number of detectors as AliCPT-1’s single band. The detailed pipeline is presented in \cite{Ghosh:2022mje}.

For the two bands of AliCPT-1, we simulate noise maps for one observational season (six months from October to March) based on a pixel-based noise covariance matrix obtained by simulating the scanning of the whole AliCPT instrument, taking into account a representative season of atmospheric emission obtained from the meteorological data MERRA-2 \cite{MERRA2}.
%The covariance matrix comes from the so-called Data Challenge 1 event\footnote{The Data Challenge 1 campaign is to build a blind data analysis with the simulated time order stream of TeSes according to the day by day scan strategy in a full observing season of AliCPT-1, in which participants attempt to estimate the power spectrum and corresponding cosmological parameters of the simulated data with the help of auxiliary data, without knowing input primordial tensor-to-scalar ratio. The data challenge allows us to test the data processing pipeline and obtain predictions of the telescope's performance.} held by the AliCPT-1 Data Analysis Group, designed to test the capability of the data simulation and data processing pipeline through blind tests.
%In the data simulation pipeline, the meteorological data MERRA-2 \cite{MERRA2} are used to estimate the atmospheric emission intensity at AliCPT-1 site, which in turn yields the corresponding amount of noise. 
199 sets of 6-month simulated noise sky maps are used to estimate the covariance matrix in the map domain.
We also use this noise covariance map to obtain the mask used for the signal sky map mentioned in the previous paragraph.
%As for the joint analysis, Planck noise maps were derived from the FFP10 simulated noise maps publicly available in the Planck Legacy Archive \cite{PLA}; while for WMAP, the noise maps were derived from realizations of the noise covariance matrix provided with the frequency maps of WMAP 9-year data \cite{WMAP9}.

The mock data in seven frequency bands is used as an input for the foreground cleaning pipeline, to obtain foreground cleaned maps. For foreground cleaning of the E-mode maps we employ the Internal Linear Combination (ILC) in the needlet domain (NILC) \citep{Delabrouille:2009, Basak:2012, Basak:2013}. The B-mode maps, with very faint signal, are processed with the constrained ILC method (cILC) in the harmonic domain \citep{Remazeilles:2010, Remazeilles:2021}.
%Extension of the ILC method to the small patch observed from the ground has been studied \cite{Zhang:2020ltv}. 

\subsection{The NILC pipeline}
An expansion of the sky maps in spherical needlets allows us to localize our statistics both in pixel domain and in harmonic domain. Assuming we have needlet bands $h_\ell^j$, where $j$ is the needlet band index and $\ell$ is the multipole, satisfying $\sum_j \left(h_\ell^j\right)^2=1$, spherical needlet functions can be defined as:
\begin{equation}
    \psi_{jk}(\hat n) = \sqrt{\frac{4\pi}{N_j}}\sum_{\ell m} h^j_\ell Y^*_{\ell m}(\hat n) Y_{\ell m}(\hat p_{jk})\,.
\end{equation}
Here we use the HEALPix pixelization, $N_j$ is the number of pixels for the $j$-th scale, 
% \textcolor{red}{(Check this! it seems to me that $N_j$ should be the number of pixels, i.e. 12\,\texttt{NSIDE}$^2$)}
and $\hat p_{jk}$ gives the spherical coordinates associated with the $k$-th pixel of the $j$-th needlet band. When a map $d(\nu, \hat n)$ is expanded in terms of $\psi_{jk}$, the coefficients of expansions are maps given by:
\begin{equation}
    b^\nu_j(\hat p_{jk}) = \int d(\nu, \hat n) \psi_{jk}(\hat n) d \hat n.
\end{equation}
For NILC we empirically estimate the cross frequency covariance for each needlet bands using the $b^\nu_j$ maps as:
\begin{equation}
    \hat C^{\nu_1\times \nu_2}_{jk} = \frac{1}{n_k} \sum_{k'} w_j(k, k') b^{\nu_1}_j(\hat p_{jk'}) b^{\nu_2}_j(\hat p_{jk'})\,,
\end{equation}
where $w_j(k, k')$ selects the domain of $n_k$ pixels around the $k$-th pixel over which we perform our averaging to estimate the covariance. The choice of $w_j(k, k')$ depends of the angular scales selected by the $j$-th needlet band. 

The NILC weights are computed as:
\begin{equation}
    W^{\rm NILC}_{\nu, j}(\hat p_{jk}) = \frac{\boldsymbol {\hat C}_{jk}^{-1} \boldsymbol a}{\boldsymbol a^t\boldsymbol {\hat C}_{jk}^{-1} \boldsymbol a}\,,
    \label{eq:nilc_wgts}
\end{equation}
where $\boldsymbol a$ is the mixing vector for the CMB. The NILC cleaned maps in needlet domain are:
\begin{equation}
    b^{\rm NILC}_j(\hat p_{jk}) = \sum_{\nu} W^{\rm NILC}_{\nu, j}(\hat p_{jk}) b^\nu_j(\hat p_{jk})\,,
\end{equation}
and finally, the cleaned CMB map is given by:
\begin{equation}
    \hat s_{\rm NILC}(\hat n) = \sum_{\ell m}\sum_{jk}b^{\rm NILC}_j(\hat p_{jk}) \sqrt{\frac{4\pi}{N_j}} h^j_\ell Y_{\ell m}(\hat p_{jk})Y_{\ell m}(\hat n)\,.
\end{equation}

\begin{figure}[t]
    \centering
    \includegraphics[width=0.6\textwidth]{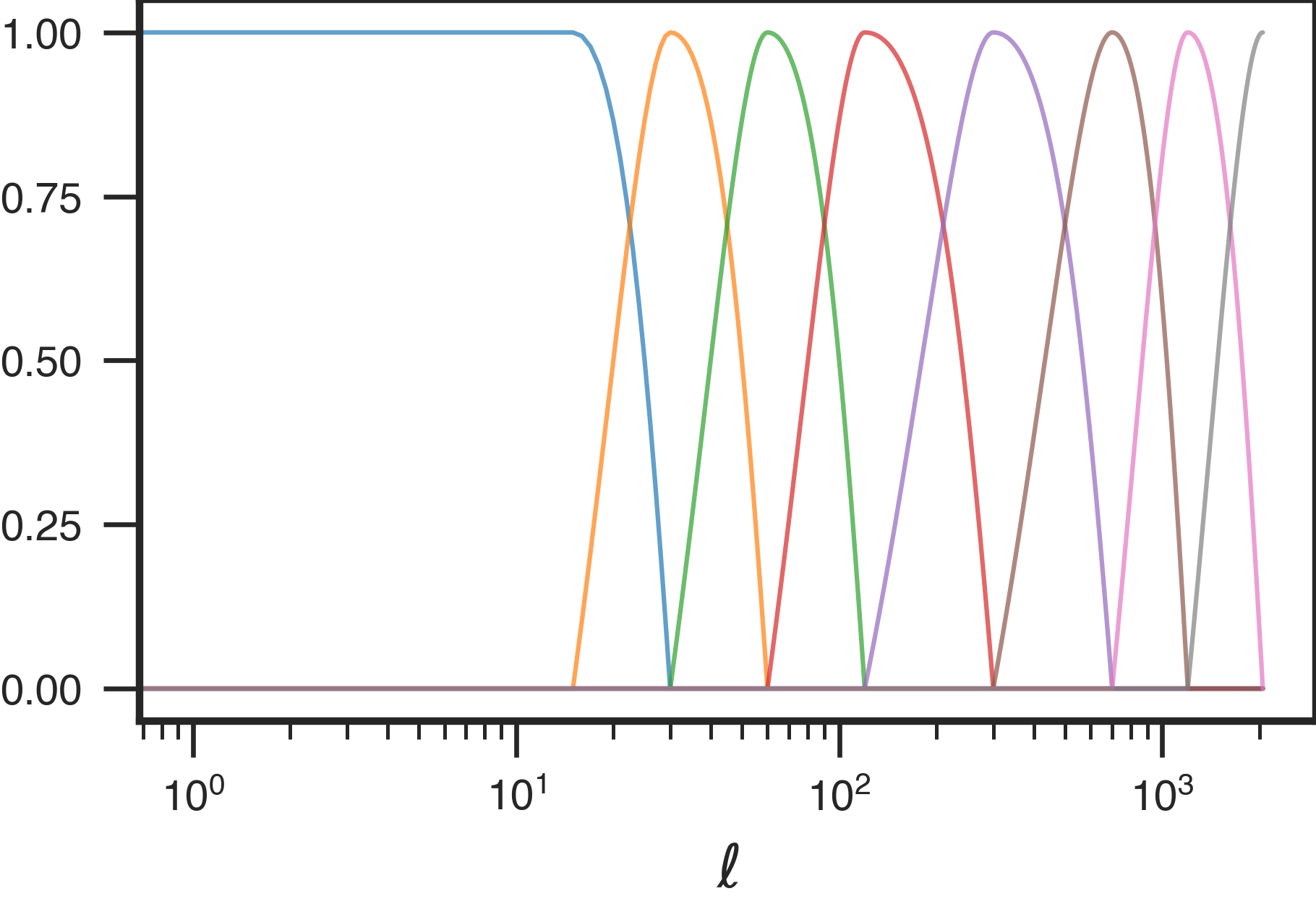}
    \caption{Cosine needlet bands used for the NILC analysis of $T$- and $E$-maps.}
    \label{fig:needlets}
\end{figure}

\begin{figure}[t]
    \centering
    \includegraphics[width=0.48\textwidth]{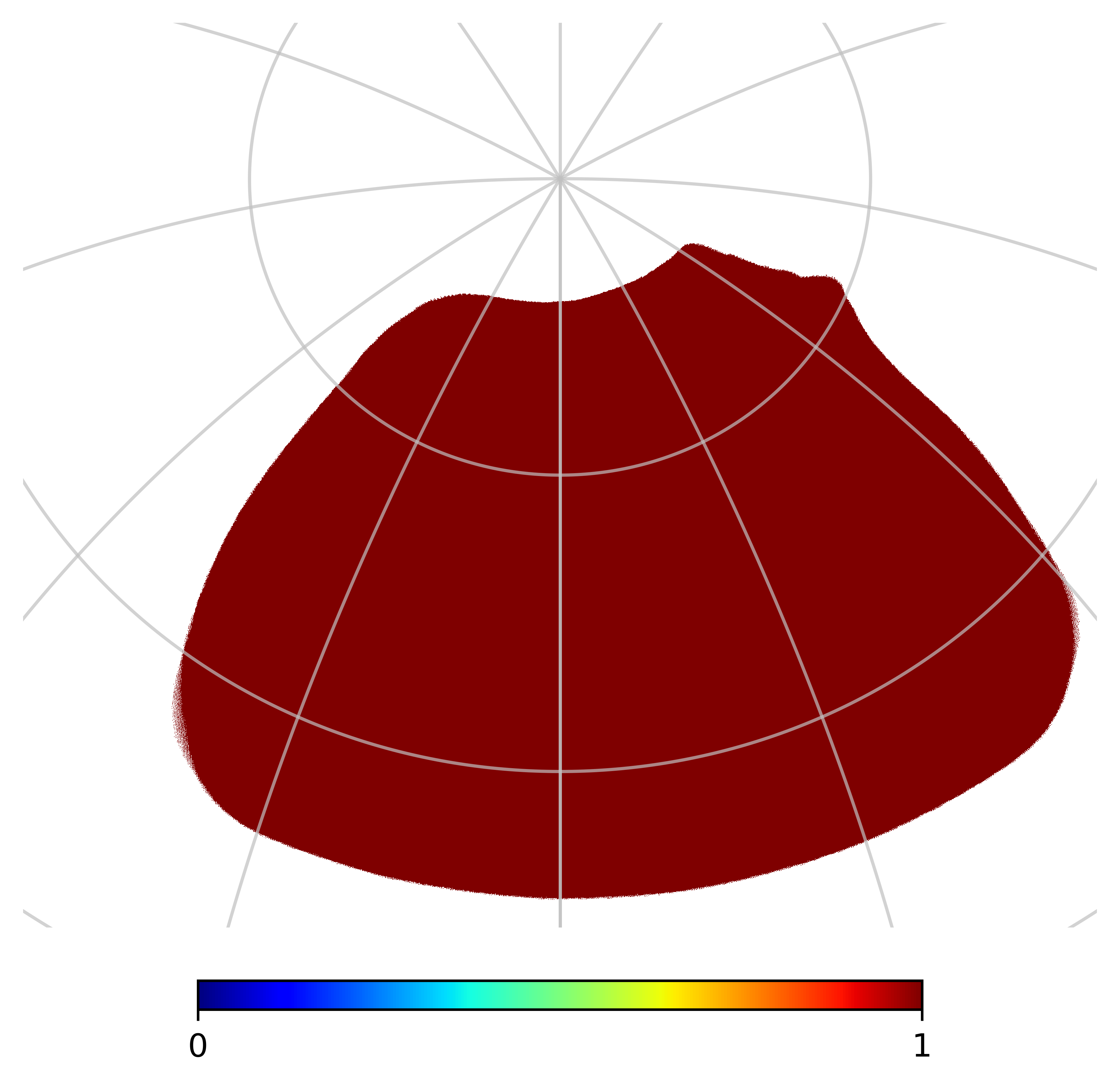}
    \includegraphics[width=0.48\textwidth]{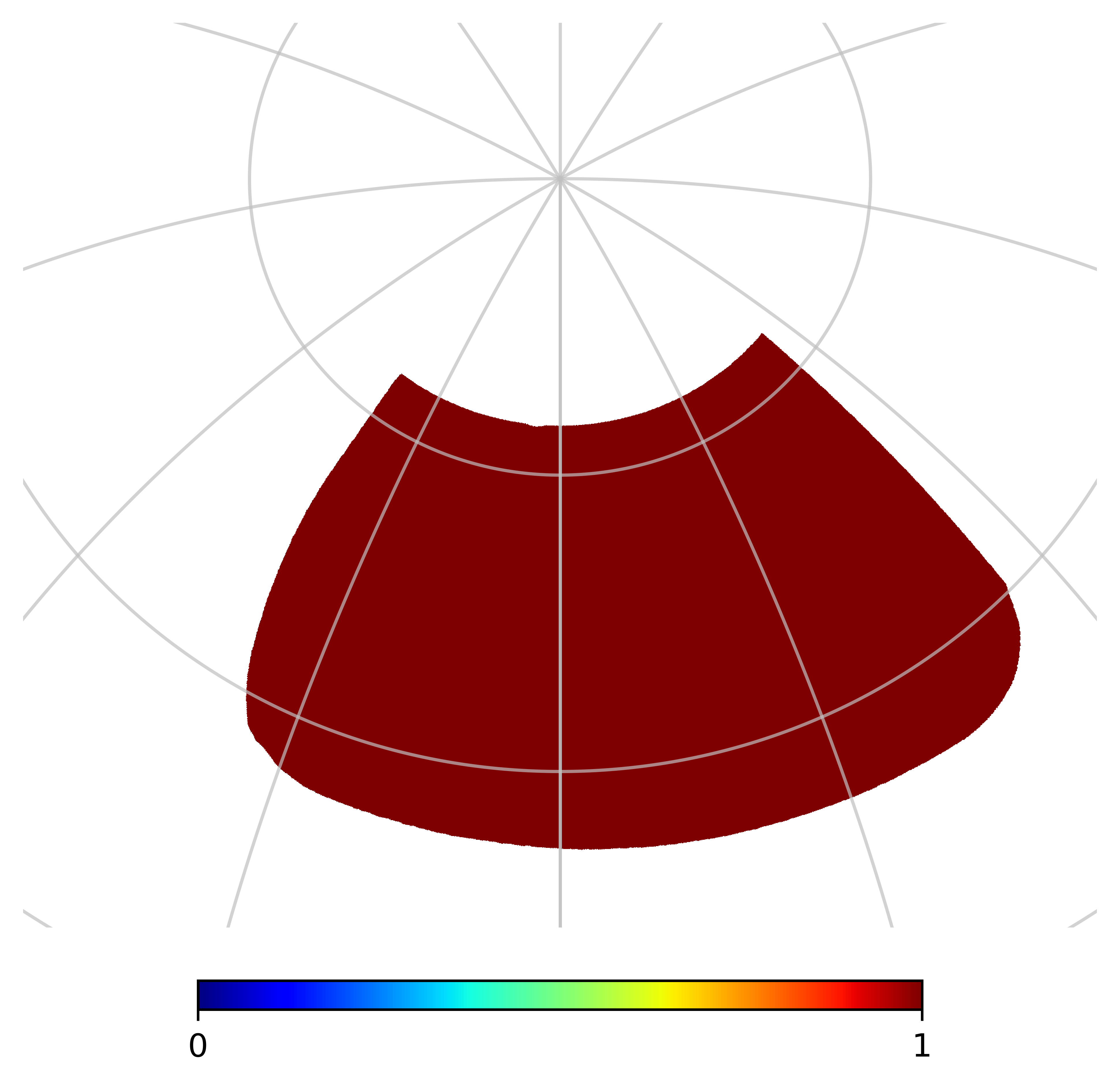}\\
    \includegraphics[width=0.48\textwidth]{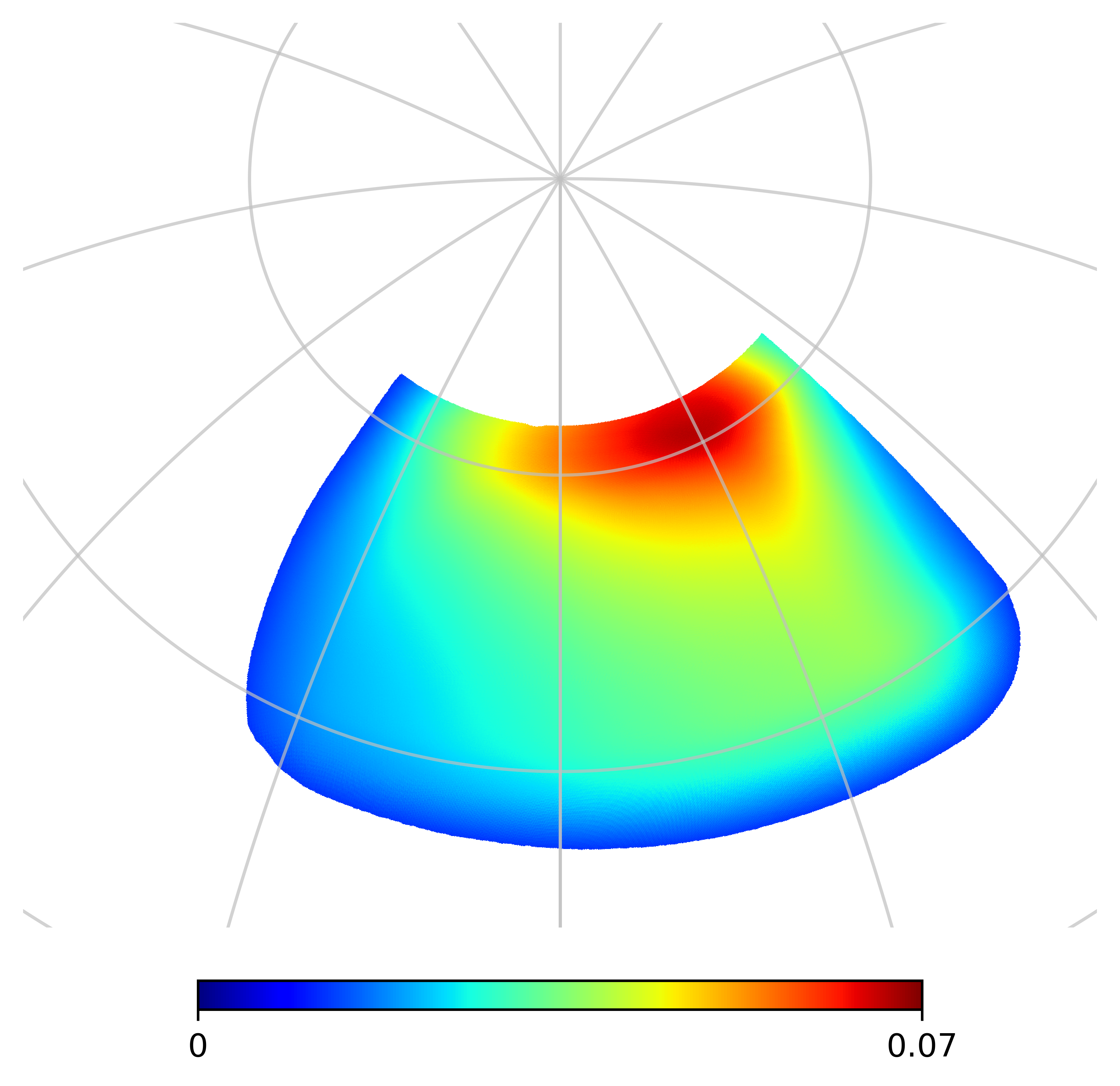}
    \caption{Various masks used in the foreground cleaning process. Top left panel is the mask used for the NILC pipelines for $T$- and $E$-maps, top right panel shows the binary mask used in the cILC pipeline for the $B$-maps. The bottom panel is the inverse AliCPT-1 noise variance weighted $B$-mode mask, used for computation of the cILC weights.}
    \label{fig:masks}
\end{figure}

The NILC implementation for this work uses 8 ``cosine-shaped'' needlet bands shown in Figure~\ref{fig:needlets}. We convolve (or deconvolve) all input maps to a common 11 arcmin angular resolution, and perform our needlet analysis with $\ell_{\rm max}=2000$. We use a mask that limits us to areas where the noise variance of the AliCPT 150 GHz channel
% \textcolor{red}{(of which channel?)} 
is smaller than 20 $\mu$K-pixel, shown in Figure~\ref{fig:masks}. The process produces foreground cleaned $E$-maps. Since the input noise and foregrounds realizations are known, we use the NILC weights to combine the individual noise inputs to compute the noise residual maps, and similarly for foreground residuals. These residual maps are useful to study the impact of residual noise and foregrounds on the lensing reconstruction results.

\subsection{The cILC pipeline}
In the cILC method, we model the $B$-mode of the polarized microwave sky in frequency $\nu$ as:
\begin{equation}
    d_\nu(\hat n) = \sum_c A_{\nu c} s_c(\hat n) + n_\nu(\hat n),
\end{equation}
where $c$ is the index over emission components, $s_c(\hat n)$ are templates of astrophysical emissions and the CMB and $n(\hat n, \nu)$ is the noise. The spatial variation of the emissions are contained in the $s_c(\hat n)$ term, while the frequency scaling of each component is captured in the mixing matrix $A_{\nu c}$. For our purpose we consider only a three component model consisting of the CMB, the polarized dust, and polarized synchrotron. The frequency scaling for dust is computed assuming a modified blackbody scaling with fixed $T_{\rm dust} =19.6$ K, and $\beta_{\rm dust}=1.59$, normalized to unity at a reference frequency of 353 GHz. For the frequency scaling of the synchrotron we assume a power law model with $\beta_{\rm sync}=-3.1$ in antenna temperature units (Kelvin RJ), and a reference frequency of 23 GHz. We account for unit correction to CMB units. The mixing vector for CMB is one at all frequencies. 

%%%%%%%%%%%%%%%%%%%%%%%%%%%%%%%%%%%%%%%%%%%
\begin{figure}[htbp]
    \centering
    \subfigure{
        \includegraphics[width=2.5in]{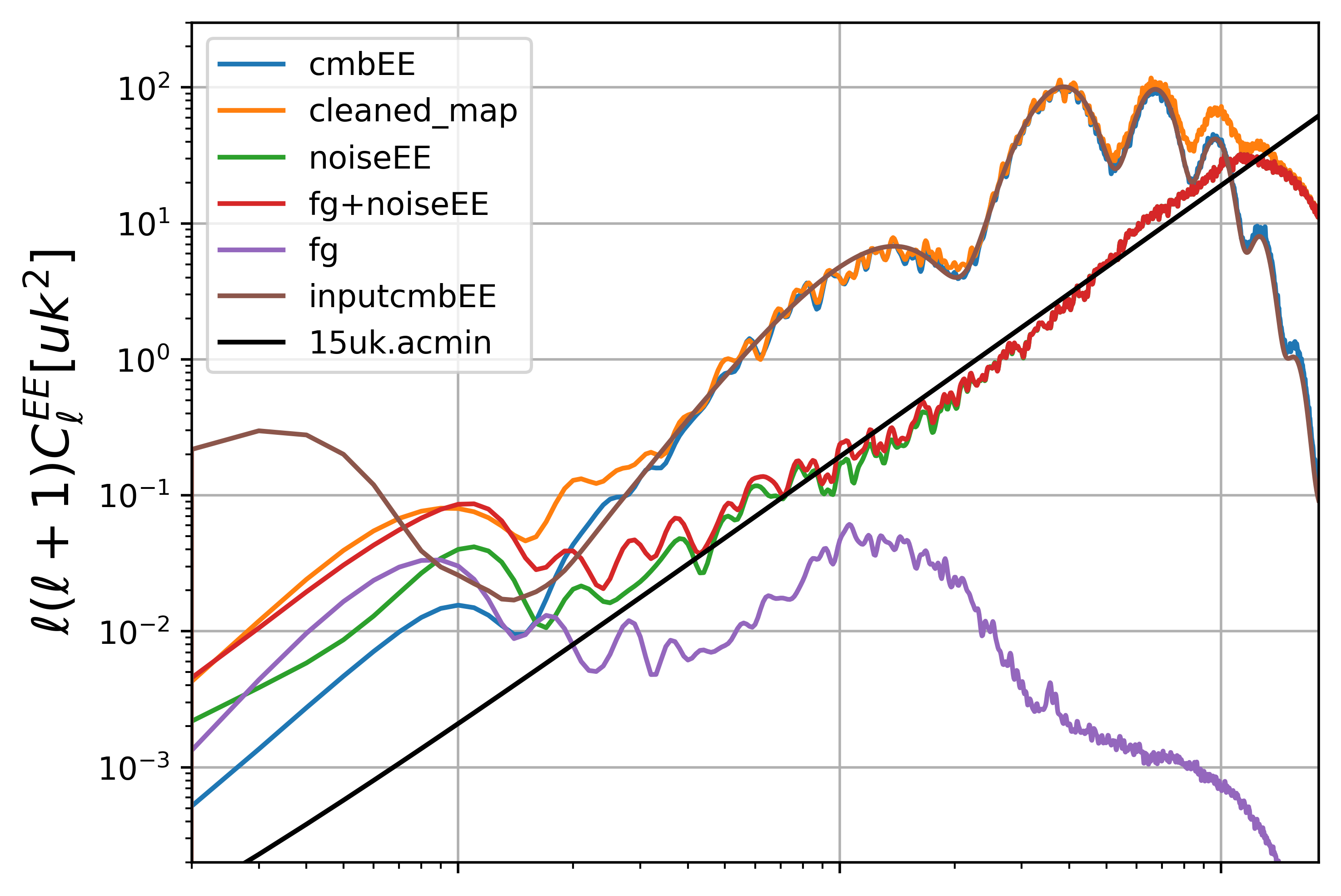}
        \label{label_for_cross_ref_1}
    }
    \subfigure{
	\includegraphics[width=2.5in]{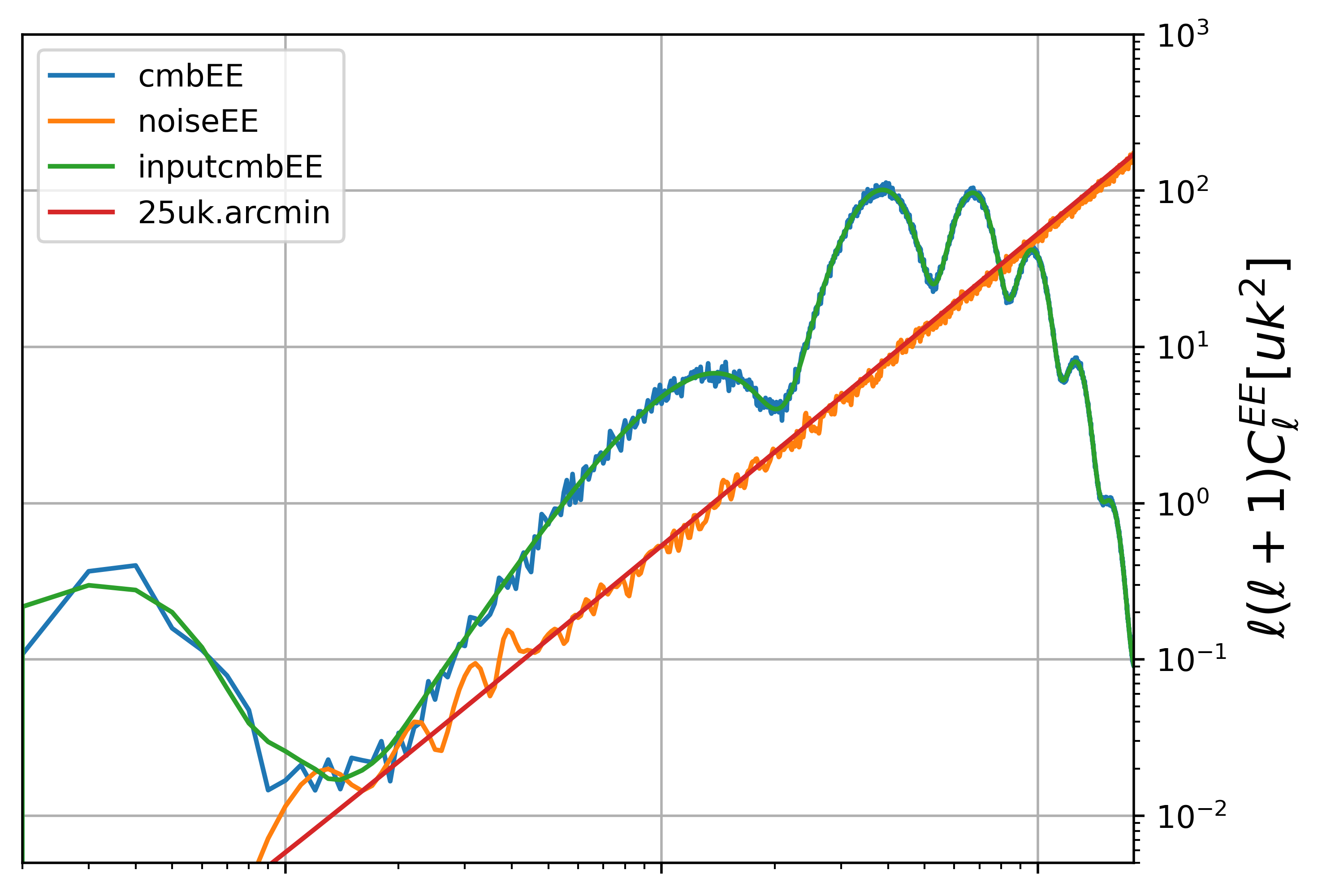}
        \label{label_for_cross_ref_2}
    }
    \quad    %用 \quad 来换行
    \subfigure{
    	\includegraphics[width=2.5in]{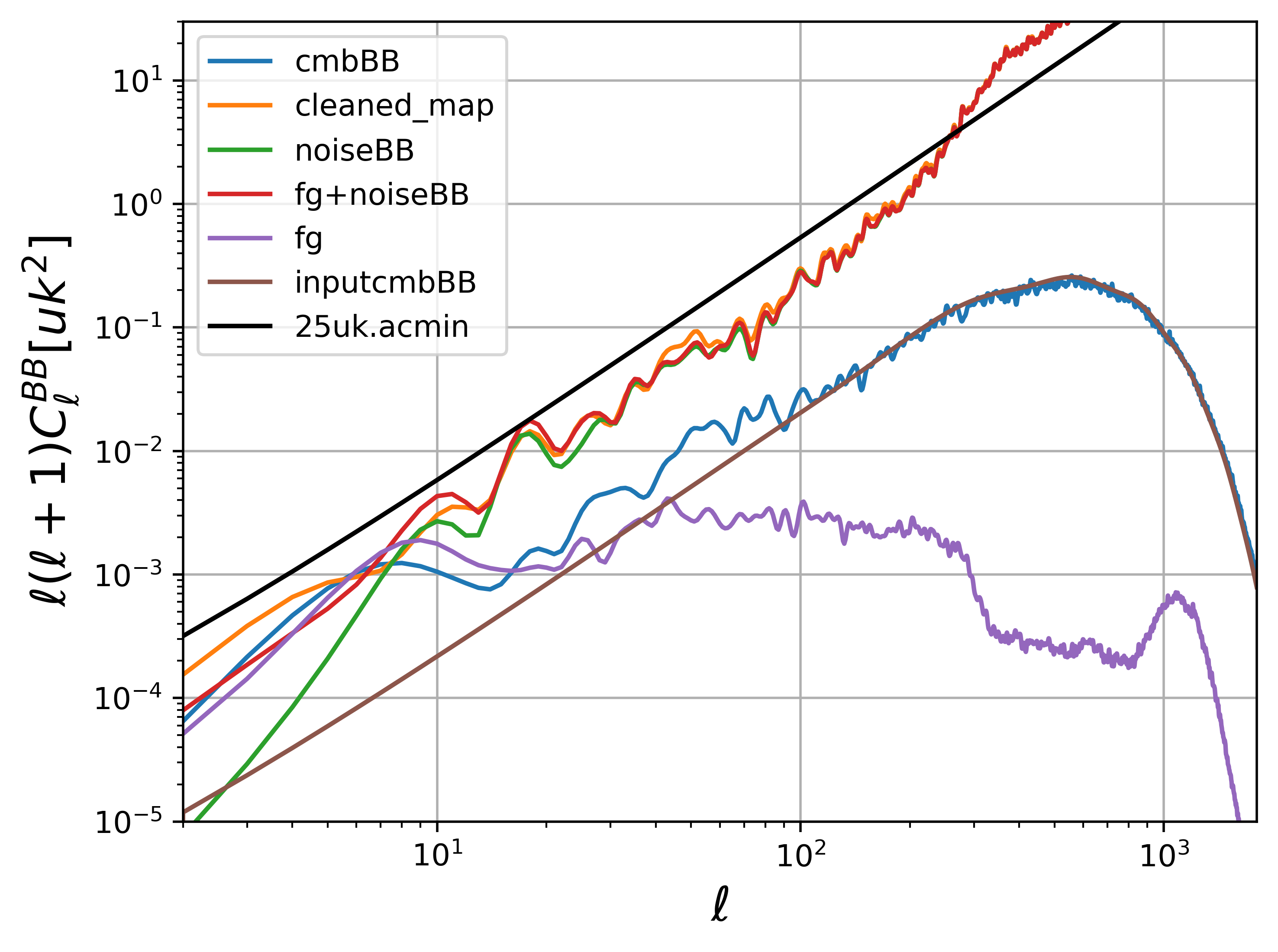}
        \label{label_for_cross_ref_3}
    }
    \subfigure{
	\includegraphics[width=2.5in]{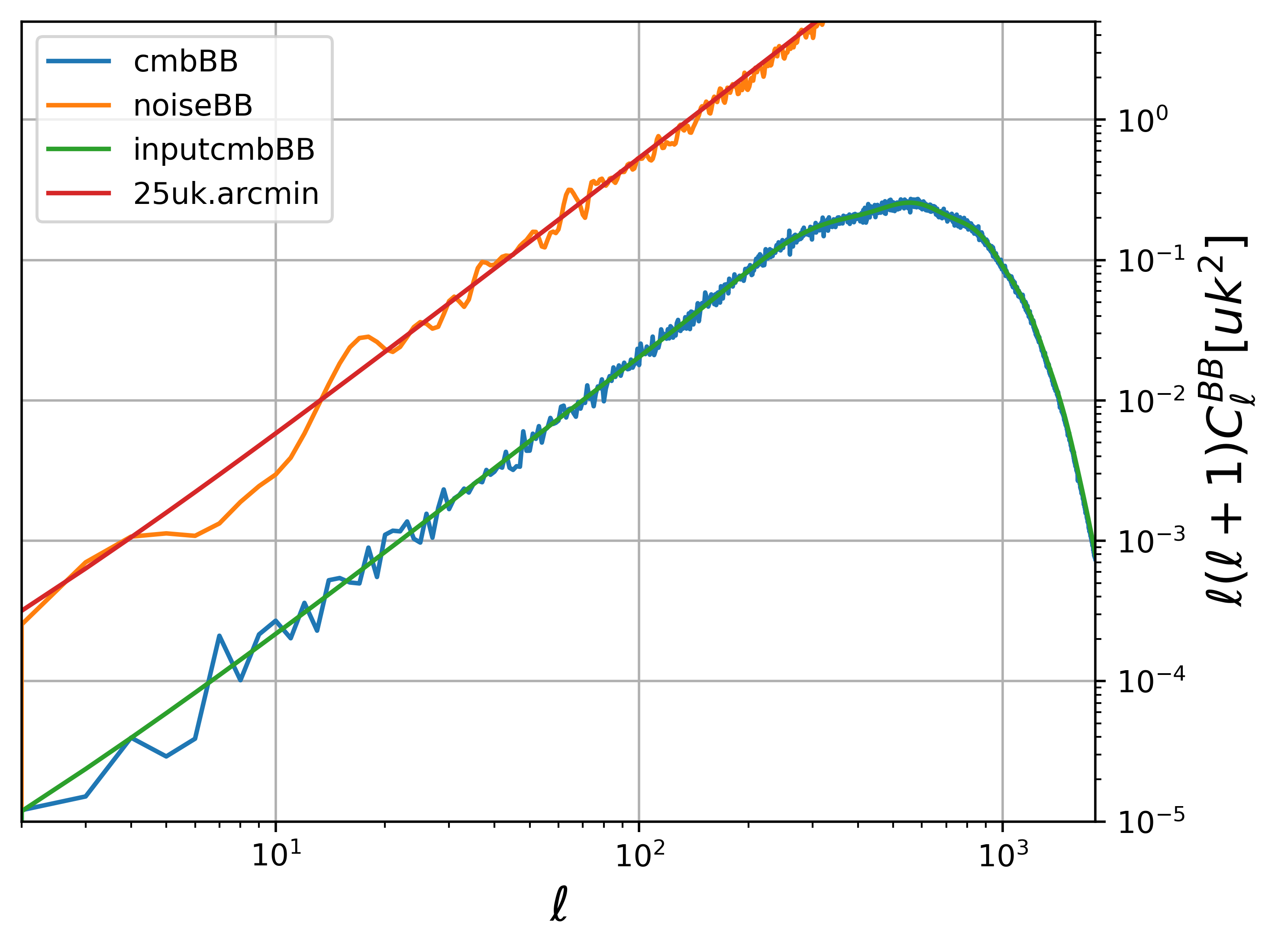}
        \label{label_for_cross_ref_4}
    }
    \caption{\label{fig:data}Power spectra of different components in the polarization map. The upper-left panel is the EE spectrum after the foreground cleaning. The lower-left panel is the same but for BB spectrum. As for comparison, we show the spectra with $25~\mu{\rm K}\cdot{\rm arcmin}$ statistical noise in the right panels, which is similar to the mock data used in Liu22 \cite{Liu:2022beb}. The foregrounds and statistical noise are denoted as `fg' and `noise', respectively. `cleaned map' denotes for `cmb+fg+noise'. In order to illustrate the effective residual noise level, we plot the $15~\mu{\rm K}\cdot{\rm arcmin}$ and $25~\mu{\rm K}\cdot{\rm arcmin}$ white noise spectra in the upper-left and lower-left panels, respectively. In BB power spectrum, we set the primordial tensor-to-scalar ratio ($r$) as 0. Hence, the BB power spectrum only shows the lensing-B mode.}
\end{figure}

The standard ILC algorithm uses one constraint, which is preserving the signal with frequency dependence given by the CMB mixing vector. In the case of the cILC implemented here, we add two additional constraints for cancelling out the average dust and synchrotron signals by using their approximate mixing vectors, assumed in the analysis to be constant over the sky area (although the actual simulations actually use varying spectral parameters). The constraints can be written as $\boldsymbol W_{\rm cILC}^t\boldsymbol A=\boldsymbol e$, where $\boldsymbol e$ is a 3 component vector with one for the CMB and zeros for the dust and synchrotron. In principle, we can add more components to capture variation of the frequency scaling of the foreground emissions. However, our data is too noisy to introduce further constraints. The cILC method obtains the weights that prioritize removing the average dust and synchrotron signal at the expense of increased variance of the cleaned map due to larger residual noise. The cILC weights are written as:
\begin{equation}
    \boldsymbol{W}^{\rm cILC} = \boldsymbol{e} \left( \boldsymbol{A}^t \boldsymbol{C}^{-1} \boldsymbol{A}\right)^{-1}\boldsymbol{A}^t \boldsymbol{C}^{-1},
\end{equation}
where $C$ is the cross frequency covariance matrix. The cILC cleaned map is obtained as $\hat s_{\rm cILC} = \sum_\nu W^{\rm cILC}_\nu d(\nu, \hat n)$. We implemented our cILC pipeline in harmonic space.

The $B$-mode maps are obtained with a mask that has smaller sky fraction. This mask further removes sky regions based on galactic foreground emissions as estimated for Planck thermal dust and synchrotron polarized intensity maps. These foregrounds and noise union mask are shown in Figure~\ref{fig:masks}. All input maps are reconvolved to an 11 arcmin beam. The $B$-mode information is dominated by the noise, even for the AliCPT-1 single season data. To estimate the covariance information correctly, the covariance matrices for the ILC weights are computed with inverse noise apodization of the input maps as shown in Figure~\ref{fig:masks}, to have greater contribution from regions of the sky with higher signal-to-noise. The ILC weights thus computed are applied on the input maps without any apodization to obtain cleaned $B$-modes maps. As for the NILC pipeline, we also compute the residual noise and foreground maps.

\subsection{Foreground cleaning efficacy and preprocessing pipeline}
After getting the foreground cleaned maps, 
%we first convert the EB maps to QU maps using the public python package \texttt{Healpy}\footnote{\url{https://healpy.readthedocs.io/en/latest/index.html}}. Then, 
we separate the results into three ingredients, the CMB signal, noise, and foreground residuals. In order to interface these foreground cleaned mock data with the lensing pipeline, which requires many realizations of the CMB itself, we generate CMB realizations with \texttt{plancklens}\footnote{\url{https://github.com/carronj/plancklens}} code. The lensed CMB maps are obtained by combining the unlensed primary CMB with the lensing potential maps via \texttt{lenspyx} code\footnote{\url{https://github.com/carronj/lenspyx}} \cite{2020ascl.soft10010C} and replace the CMB signal in the ILC results with those realizations. We save the corresponding lensing potential for each realization for comparison. 
Figure \ref{fig:data} shows the signal, foreground residual and statistical noise power spectra of EE and BB. 
The left panels show that the foreground residuals in the polarization data is significantly lower than the statistical noise. For E-mode (top-left panel), the typical statistical noise is about $15\mu{\rm K}\cdot{\rm arcmin}$. The statistical noise in the B-mode (bottom-left panel) slightly deviates from white noise shape ($\sim\ell^2$), but is of the order of $25\mu{\rm K}\cdot{\rm arcmin}$ level. The main reason for different statistical noise level between E- and B-maps is the use of different foreground removal algorithms. The foreground cleaning method adopted in this paper is also used for the primordial gravitational wave search in B-map, which requests more careful removal of the foregrounds than in the E-map case.
%For E-map, we used the needlet Internal Linear Combination (NILC). 
%For B-map, we employ the constrained ILC method (cILC) in the harmonic domain, which utilizes additional constraints on average frequency scaling of the dust and synchrotron to remove these at the expense of noisy maps.
For comparison, we show in the right panels of Fig. \ref{fig:data} the mock data, which has similar noise level as those used in Liu22 \cite{Liu:2022beb}. One can see that the typical statistical noise is about $25\mu{\rm K}\cdot{\rm arcmin}$ (red line in the right panels). 

%%%%%%%%%%%%%%%%%%%%%%%%%%%%%%%%%%%%%%%%%%%
\section{Lensing reconstruction results}
\label{sec:3}
We follow the lensing reconstruction formalism presented in the Planck 2018 lensing paper \cite{Aghanim:2018oex}, which calculates the lensing estimator from pairs of filtered maps \cite{Hu:2001kj,Okamoto:2003zw,Maniyar:2021msb,Carron:2017mqf}. One leg of the pair is Wiener-filtered map and the other leg is filtered with inverse variance. The variance in the filter is derived from the mock simulations. The inversion of the covariance matrix is computed via a conjugate-gradient inversion method with a multi-grid preconditioner~\cite{Smith:2007rg}, and contains both the foreground residual and the statistical noise. After the filtering, we combine the pair maps into the quadratic form according to the equations (3.4)-(3.8) listed in \cite{Liu:2022beb}. Due to the presence of mask and foreground residuals, there will be extra statistical anisotropies in the map even without any lensing signals. 
They will bias the estimation of lensing potential. 
In order to remove this bias, we first calculate their contributions via Monte Carlo simulations. Then, we compute the average value (mean-field) of the quadratic estimator. This average is a representation of the contribution from the extra statistical anisotropy source, which is subtracted from the original estimator for mean-field subtraction. 
In this work, we calculate the mean-field using 44 sets of simulations. 

\begin{figure}[tbp]
\centering 

\includegraphics[width=0.47\textwidth]{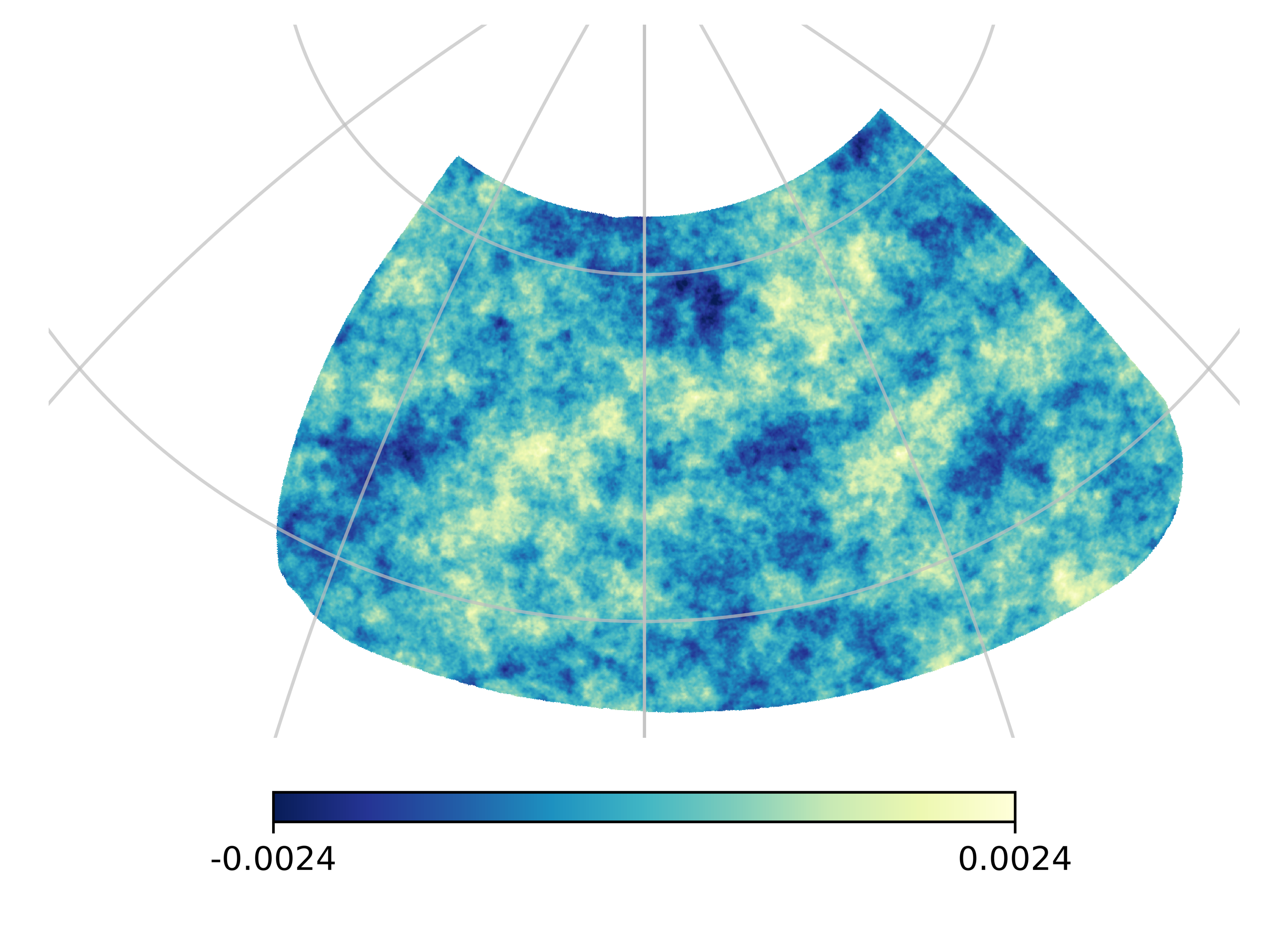}
\includegraphics[width=0.47\textwidth]{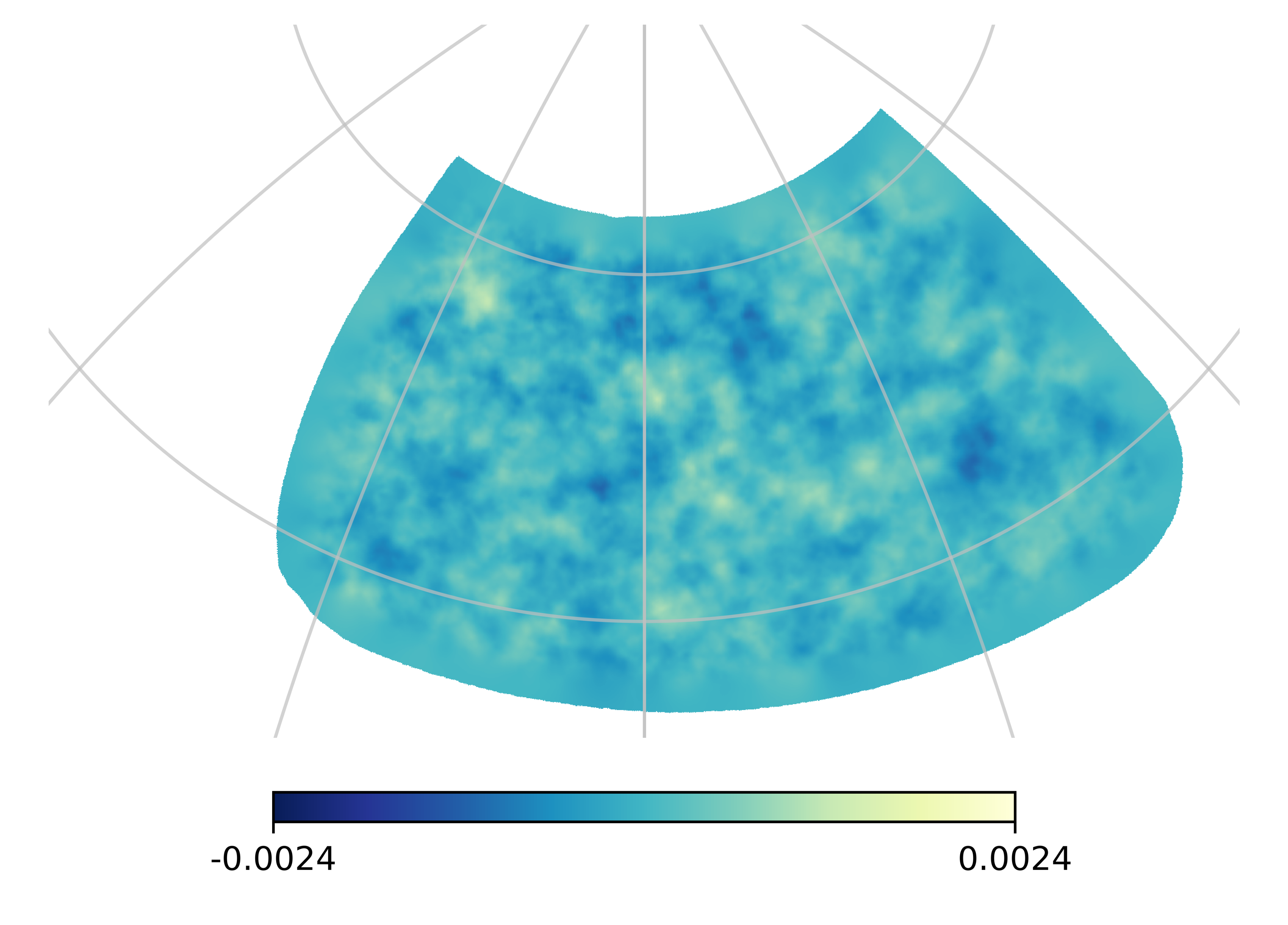}

\caption{\label{fig:i}Reconstructed lensing map. The left panel is the input lensing map; the right panel is the reconstructed map from the polarization estimator. In order to highlight the lens structures, we plot the Wiener-filtered deflection angle amplitude $\hat\alpha^{\rm WF}=\sqrt{L(L+1)}\hat\phi_{LM}C_L^{\phi\phi,{\rm fid}}/[C_L^{\phi\phi,{\rm fid}}+N_L^{(0),{\rm ana}}]$.}
\label{fig:noise_var}
\end{figure}

The normalization factor of the lensing field is calculated in two steps. Firstly, by evaluating the averaged noise level over the whole patch, we calculate the normalization factor according to analytical formula with an evaluated isotropic noise level. Then, we calibrate normalization bias originated from the noise inhomogeneities effect via the numerical simulation. After this, we get the reconstructed lensing map, as shown in Figure \ref{fig:noise_var}. In order to highlight the lens structures, we plot the Wiener-filtered deflection angle amplitude $\hat\alpha^{\rm WF}=\sqrt{L(L+1)}\hat\phi_{LM}C_L^{\phi\phi,{\rm fid}}/[C_L^{\phi\phi,{\rm fid}}+N_L^{(0),{\rm ana}}]$.
The left panel is the input data and the right one is the reconstructed deflection angle from the polarization estimator. 
One can see that due to the sub-optimal noise performance, the reconstruction can only capture a few features, such as the dark blue spots in the lower-right corner in Figure \ref{fig:noise_var}.

%%%%%%%%%%%%%%%%%%%%%%%%%%%%%%%%%%%%%%%%%%%

According to these reconstructed lensing maps, the raw power spectrum of the lensing potential is simply the quadrature of multipoles with the same $L$ but different $M$s
\begin{equation}
    \label{eq:phihatpow}
    \hat{C}_L^{\hat\phi\hat\phi}=\frac{1}{(2L+1)f_{\rm sky}}\sum_{M=-L}^{L}\hat\phi_{LM}\hat\phi_{LM}^{\ast}\;,
\end{equation}
where $\hat\phi_{LM}$ are the harmonic transformation of the reconstructed lensing potential. The quadratic estimator spectrum contains not only the sought-after signal, but also unavoidably the Gaussian reconstruction noise sourced by the CMB and instrumental noise ($N_0$ bias) and the non-primary couplings of the connected 4-point function \cite{Kesden:2003cc} ($N_1$ bias). After subtracting these biases, we obtain the final estimated power spectrum
\begin{equation}
    \label{eq:phipow}
    \hat C^{\phi\phi}_L=\hat C_L^{\hat\phi\hat\phi}-\Delta C_L^{\hat\phi\hat\phi}|_{\rm RDN0}-\Delta C_L^{\hat\phi\hat\phi}|_{\rm N1}\;.
\end{equation}
``RDN0'' means realization-dependent $N_0$ bias, which is designed to subtract the primary CMB contamination in the most faithful manner.
For further detailed expressions, we refer to the Planck 2013/2015 lensing papers \cite{Ade:2013tyw,Ade:2015zua}.

\begin{figure}[tbp]
\centering 
\includegraphics[width=0.9\textwidth]{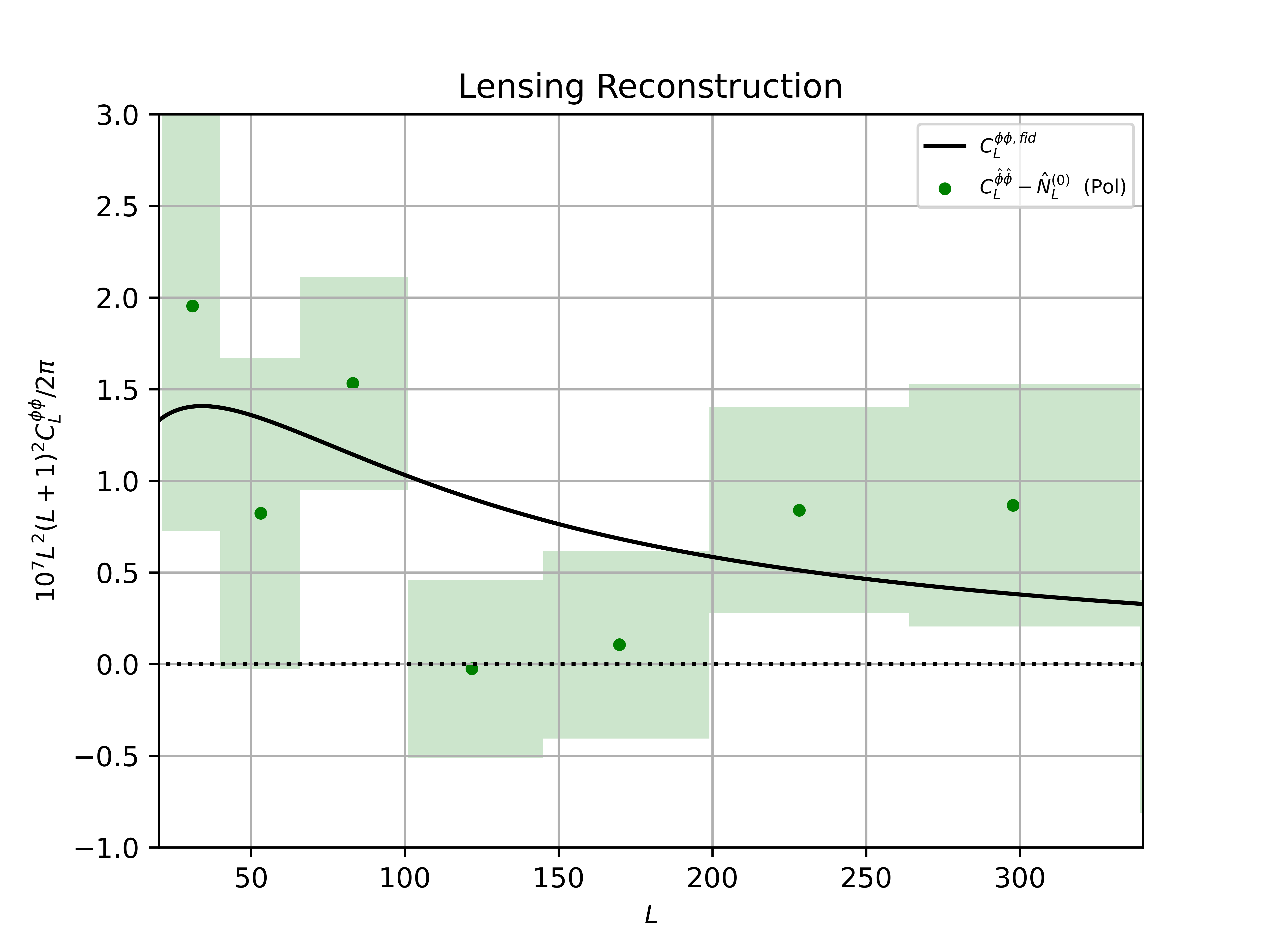}
\caption{\label{fig:recon}Reconstructed lensing potential power spectrum from polarization estimator (green points). The colored boxes denote for the 1$\sigma$ error regions calculated from 154 sets of simulations.}
\end{figure}

\begin{figure}[tbp]
\centering 
\includegraphics[width=0.9\textwidth]{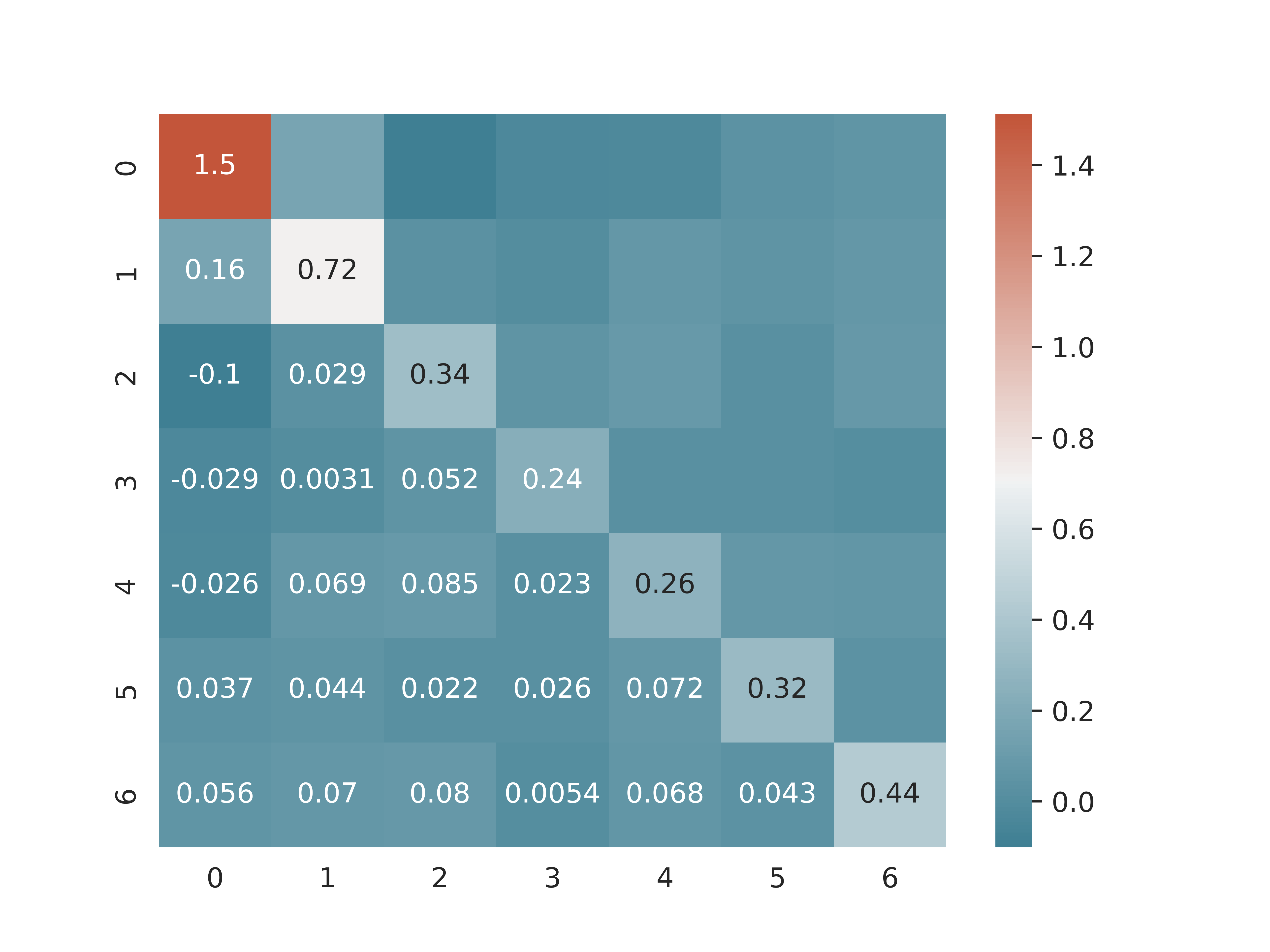}
\caption{\label{fig:cov}Covariance of convergence power spectrum. The row and column index are the index of the $\ell$-bins. The numbers in the matrix element are the covariance values.}
\end{figure}

In Figure \ref{fig:recon}, we show the reconstructed lensing potential power spectrum from polarization data, which combining EE, EB, and BB estimators. 
Here, we plot the multipole range from $20$ to $340$, which contributes the main part of SNR.
There are seven $\ell$-bins, in two of them the reconstructed value deviate from the theoretical prediction (black curve) about $1\sigma$ level. 
From the simple Gaussian statistics by assuming each $\ell$-bin are independent, this result is fairly normal. 
To further affirm the above intuition, we show the covariance in Figure \ref{fig:cov}.  
One can see that the diagonal term is roughly one order magnitude higher than the off-diagonal one.
Indeed, one can further figure out the origins of these deviations from the EE noise power spectrum.
As shown in the upper-left panel of Figure \ref{fig:data}, there is a bump in the foreground residue (pure curve) in the range of $\ell\in(100,200)$. 
We think this feature is responsible for the deviation from the standard prediction. 

%%%%%%%%%%%%%%%%%%%%%%%%%%%%%%%%%%%%%%%%%%%

Finally, we calculate the lensing power spectrum detection SNR via the Fisher matrix method
\begin{equation}
    \label{eq:SNR}
    {\rm SNR}=\sqrt{\sum_{\ell,\ell'}C_{\ell}\mathbb{C}^{-1}_{\ell\ell'}C_{\ell'}}\;,
\end{equation}
where the numerator $C_{\ell}$ is the theoretical lensing potential spectrum, and the $\mathbb{C}_{\ell\ell'}$ is the covariance matrix obtained from our data sets via
\begin{equation}
\label{eq:covqu}
    \mathbb{C}_{\ell\ell'} = \frac {1}{N-1} \sum\limits_{n=1}^{N=154} \bigg[\Big(\hat C^{\kappa\kappa}_{\ell} - \overline{C}^{\kappa\kappa}_{\ell}\Big)\times \Big(\hat C^{\kappa\kappa}_{\ell'} - \overline{C}^{\kappa\kappa}_{\ell'}\Big) \bigg] \ ,
\end{equation}
in which $\overline{C}^{\kappa\kappa}_{\ell}$ is the averaged convergence power spectrum across the data sets.
The value of the covariance is shown in Figure \ref{fig:cov}. 
The final SNR obtained from multipole range of $\ell\in(20,1000)$ is about SNR$\approx4.5$, in which multipoles of $\ell\in(20,340)$ contribute the major part, namely SNR$\approx4.1$.   
As a consistency check, the polarization estimator without including the foregrounds from Liu22 \cite{Liu:2022beb} reports an SNR$\approx6.6$.  
This lensing reconstruction capability is similar as other stage-III small aperture millimeter telescope, such as BICEP2+BICEP3+{\it Keck~Array} data up to the 2018 seasons give $A_L^{\phi\phi}=0.95\pm0.20$ \cite{BICEPKeck:2022kci}.

%%%%%%%%%%%%%%%%%%%%%%%%%%%%%%%%%%%%%%%%%%%

%%%%%%%%%%%%%%%%%%%%%%%%%%%%%%%%%%%%%%%%%%%
\section{Conclusion}
\label{sec:4}

In this paper, we update the AliCPT-1 lensing reconstruction analysis by adding the foreground contamination. 
In the previous study, we only consider the statistical noise according to a given scan strategy. 
We used the state-of-the-art extra-galactic foreground model in the millimeter band, the Planck Sky Model (PSM).  
Our simulations include galactic thermal dust, synchrotron, free-free, spinning dust, CO emission, and extragalactic foreground emissions. We perform foreground cleaning with a combination of the commonly used Needlet Internal Linear Combination (NILC) and the constrained Internal Linear Combination (cILC).
With 4 modules 1 year observation data in 7 bands including: 90 and 150GHz dual bands of AliCPT-1, 100GHz, 143GHz, 217GHz, 353GHz four HFI bands of Planck, WMAP-K band (23GHz)we get a residual noise in E- and B-maps of roughly $15~\mu{\rm K}\cdot{\rm arcmin}$ and $25~\mu{\rm K}\cdot{\rm arcmin}$, respectively. 
Thanks to the good performance of the foreground cleaning operation, the foreground residual noise is sub-dominant. The final CMB lensing reconstruction signal-to-noise ratio (SNR) from polarization data is about SNR$\approx4.5$. 
This lensing reconstruction ability is similar to what we can get from the latest BICEP2+BICEP3+{\it Keck~Array} data \cite{BICEPKeck:2022kci}, which have better noise performance (effectively $2.8~\mu{\rm K}\cdot{\rm arcmin}$) but worse spatial resolution ($43$ arcmin FWHM in 95GHz/$30$ arcmin FWHM in 150GHz/$20$ arcmin FWHM in 220GHz) \cite{BICEP2:2019upn}. 
Unlike Liu22 \cite{Liu:2022beb}, the numbers presented in this work are based only on the simulation data by assuming ``$\texttt{4 module*yr}$'' configuration. As demonstrated in Liu22, additional observing time  in the final ``$\texttt{48 module*yr}$'' data significantly improves the lensing reconstruction results. %However, this will ask for much longer simulated data, which is currently un-available. Hence, we will 
We leave to future work the study of the improvement of lensing reconstruction with more observing time in the presence of foreground residuals. 

%%%%%%%%%%%%%%%%%%%%%%%%%%%%%%
\acknowledgments

This work is supported by the National Key R\&D Program of China No. 2020YFC2201603.

%\paragraph{Note added.} This is also a good position for notes added
%after the paper has been written.


\end{document}